\documentclass{PoS}

\usepackage{lineno}
\usepackage[htt]{hyphenat}
\usepackage{multirow}
\usepackage{subfigure}
\usepackage{amsmath}
\usepackage{amssymb}
\usepackage{url}
\usepackage{color}
\usepackage[normalem]{ulem}

\DeclareUrlCommand\ULurl{%
  \renewcommand\UrlLeft{\uline\bgroup}%
  \renewcommand\UrlRight{\egroup}}

\title{Performance of Monte Carlo Event Generators for the Production of Boson and Multi-Boson States ATLAS Analysis}

\ShortTitle{Monte Carlo Event Generators for Boson and Multi-Boson States in ATLAS}

\author{\speaker{Francesco Giuli}\thanks{on behalf of the ATLAS Collaboration}\\
        University of Oxford\\
        E-mail: \email{francesco.giuli@cern.ch}, \email{francesco.giuli@physics.ox.ac.uk}}

\abstract{The Monte Carlo (MC) setups used by ATLAS to model boson$+\mathrm{jets}$ and multi-boson processes at $\sqrt{s}$ = 13 TeV in proton-proton collisions are described. Comparisons between data and several event generators are provided for key kinematic distributions. Issues associated with the evaluation of systematic uncertainties are also discussed. This proceeding is a summary of the results collected in two recent ATLAS PUB notes published for the MC workshop jointly-organised by the the ATLAS and CMS Collaborations and held at CERN in May 2017.}

\FullConference{EPS-HEP 2017, European Physical Society conference on High Energy Physics\\
		5-12 July 2017\\
		Venice, Italy}

\begin{document}

\tableofcontents

\paragraph{Introduction}
This proceeding describes the Monte Carlo (MC) used by ATLAS~\cite{ATLAS} to model boson$+\mathrm{jets}$ ($V+\mathrm{jets}$)~\cite{boson} and multi-boson ($VV/VVV$)~\cite{multiboson} processes in 13 TeV $pp$ collisions. The baseline MC generators are compared with each other in key kinematic distributions of the processes under study. Sample normalisation and assignment of systematic uncertainties are discussed.

\begin{figure}[t]
\centering
\subfigure[]{\includegraphics[width=.48\textwidth]{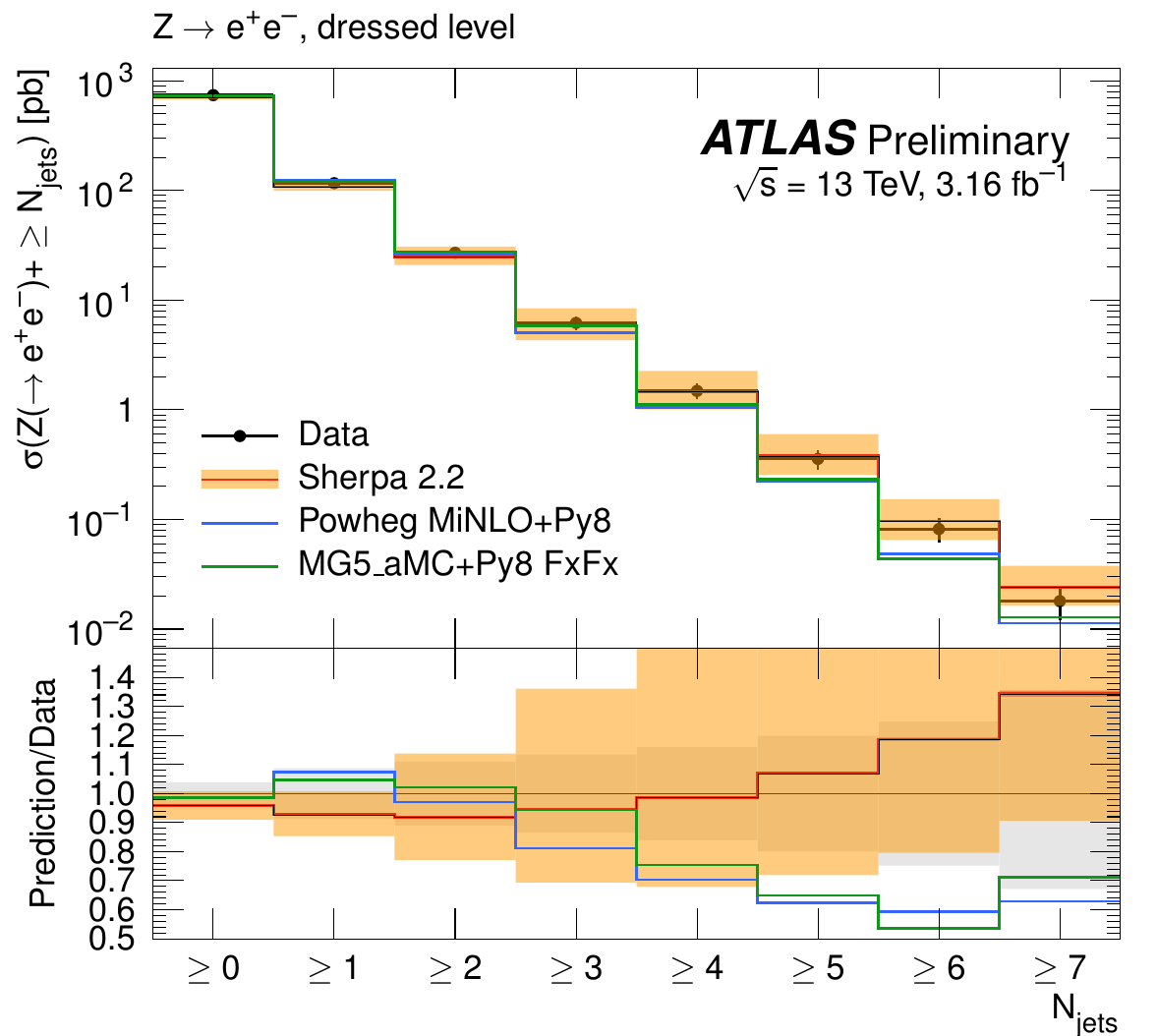}}
\subfigure[]{\includegraphics[width=.48\textwidth]{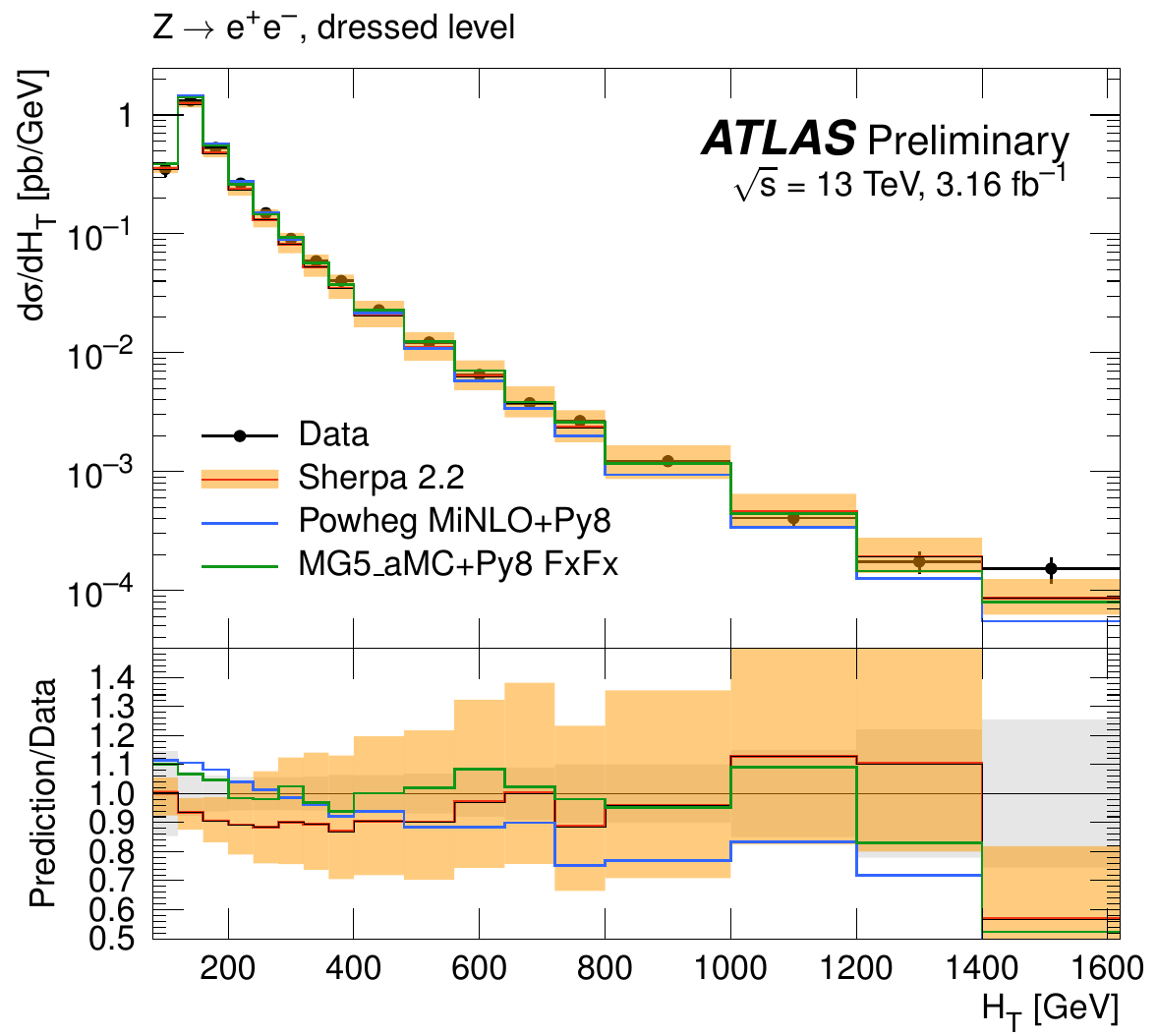}} 
\caption{Predictions for the differential cross sections as a function of the inclusive jet multiplicity (left) and of the scalar jet-$p_\mathrm{T}$ sum, $H_\mathrm{T}$, (right) from {\tt{Sherpa 2.2}} (red), {\tt{Powheg MiNLO+Pythia8}} (blue) and {\tt{MG5{\_}aMC@NLO+Py8}} using FxFx (green). The predictions are compared to data (black) from a recent ATLAS measurement of $Z+\mathrm{jets}$ production~\cite{Zjets}. These plots are taken from Ref.~\cite{boson}.}
\label{Zjets}
\end{figure}

\begin{figure}[t]
\centering
\subfigure[]{\includegraphics[width=.48\textwidth]{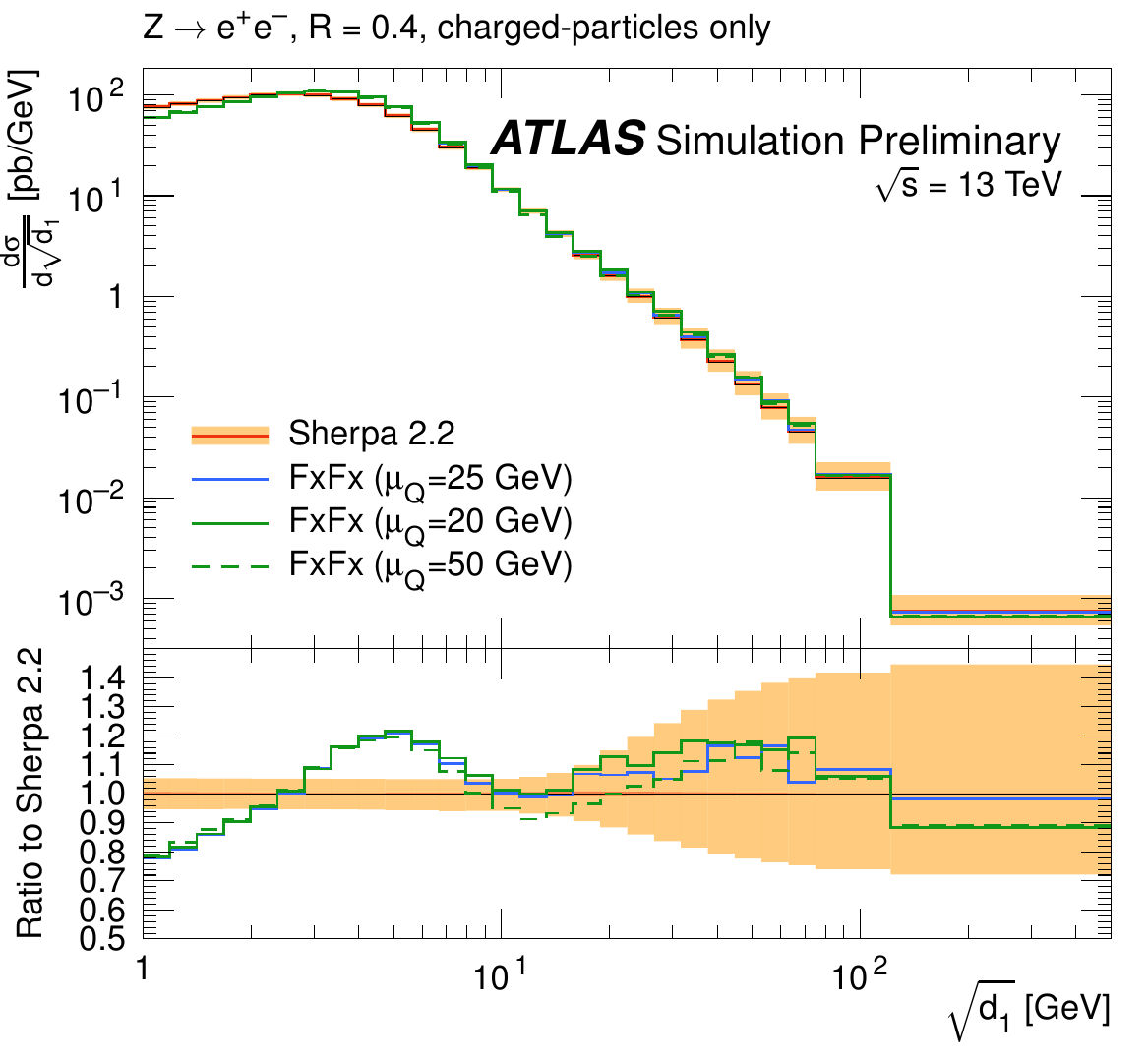}}
\subfigure[]{\includegraphics[width=.48\textwidth]{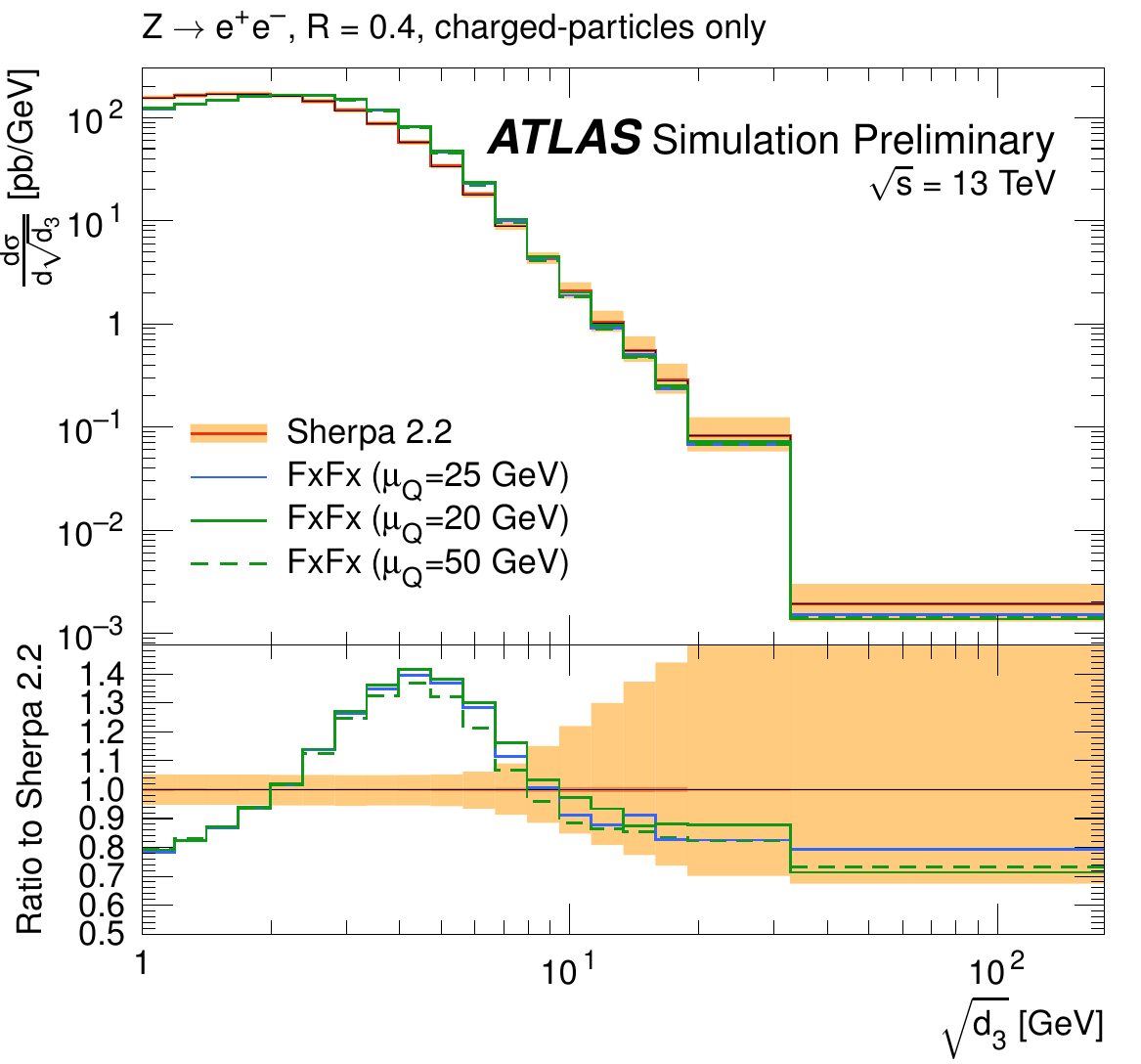}}
\caption{Predictions for the differential cross sections as a function of first-order splitting scale, $\sqrt{d_{1}}$ (left), and of the third-order splitting scale, $\sqrt{d_{3}}$ (right), in the $k_{t}$ algorithm using a jet-radius parameter of $R$ = 0.4~\cite{kTscales} from {\tt{Sherpa 2.2}} (red) and {\tt{MG5{\_}aMC@NLO}} using FxFx using $\mu_{Q}$ of 20 GeV (blue), 25 GeV (solid green) and 50 GeV (dashed green). These plots are taken from Ref.~\cite{boson}.}
\label{kTscales}
\end{figure}

\paragraph{Boson+jets}
The setup of the samples used for the different comparisons is described in the following:
\begin{itemize}
\item {\bf{Sherpa}} {\tt{Sherpa}}~\cite{SherpaMC} is a parton shower MC generator simulating additional hard parton emissions that are matched to a parton shower based on Catani-Seymour subtraction terms~\cite{Catani}. The merging of multi-parton matrix elements (ME) with the parton shower (PS) is achieved using an improved CKKW matching procedure~\cite{Hoche,CataniQCD}, which is extended to next-to-leading order (NLO) accuracy using the MEPS@NLO prescription~\cite{HocheQCD}.
\item {\bf{MadGraph5{\_}aMC@NLO} using CKKW-L} Matrix elements for $V$+ up to 4 partons at LO accuracy are produced using {\tt{MadGraph5{\_}aMC@NLO v2.2.2}}~\cite{Alwall} interfaced with {\tt{Pythia v8.186}}~\cite{Pythia8} for the modelling of the parton shower and underlying event. The CKKW-L matching~\cite{Hoche,ckkwl} and merging procedure is applied, with a merging scale ($\mu_{Q}$) of 30 GeV. The NNPDF3.0NLO PDF set is used with $\alpha_{s}$ = 0.118 and the A14 tune of Pythia8~\cite{A14} is applied.
\item {\bf{MadGraph5{\_}aMC@NLO} using FxFx} Samples have also been generated using the {\tt {MadGraph5{\_}aMC@NLO}} program to generate matrix elements for $V$ + 0, 1 and 2 partons to NLO accuracy. The showering and subsequent hadronisation has been performed using Pythia 8.210 with the A14 tune, using the NNPDF2.3LO PDF set with $\alpha_{s}$ = 0.13. The different jet multiplicities are merged using the FxFx prescription~\cite{FxFx} implemented in the {\tt{MadGraph5{\_}aMC@NLO}} program (version 2.3.3 is used here). The impact of various $\mu_{Q}$ has been studied, analysing three different values: 20 GeV (downward variation), 25 GeV (nominal value) and 50 GeV (upward variation). 
\item {\bf{Powheg MINLO}} Predictions from {\tt {Powheg MiNLO}}~\cite{Alioli,Minlo,Hamilton} interfaced to Pythia 8.210\\
~\cite{Pythia8} with the AZNLO tune~\cite{AZNLO} were obtained to produce $V$ + jets events. The PDF set used in Powheg is CT14NNLO~\cite{CT14} whereas the PDF set used in the parton shower is the CTEQ6L1~\cite{CTEQ6} LO set.
\end{itemize}
Figure~\ref{Zjets} shows comparisons of {\tt{Sherpa 2.2}} (red), {\tt{Powheg MiNLO+Py8}} (blue) and {\tt{MG5{\_}aMC@NLO+Py8}} using FxFx (green) against data (black) from the 13 TeV ATLAS $Z+\mathrm{jets}$ measurement~\cite{Zjets}. The measurement uncertainty is indicated by a grey band in the ratio panels, while generator uncertainties have been estimated using Sherpa 2.2 and are indicated by the orange band in the ratio panels. The MEPS@NLO setup using Sherpa 2.2 is NLO-accurate for up to two extra emissions and LO-accurate for up to four extra emissions. At very high jet multiplicities, Sherpa 2.2 starts to diverge from the data, indicating too much activity in the PS. The {\tt{MG5{\_}aMC@NLO}} setup already begins to mismodel the data for lower multiplicities, as it lacks additional multilegs beyond the third emission, making it less attractive for many new-physics searches interested in a good modelling of high multijet configurations, even though the mismodelling is likely covered by the scale uncertainties. 
A similar comparison for the first- and third-order splitting scale distribution occuring in the $k_{t}$ algorithm~\cite{CataniKT,SalamKT} is shown in Figure~\ref{kTscales} between {\tt{Sherpa 2.2}} (red) and {\tt{MG5{\_}aMC@NLO+Py8}} using FxFx (blue) at a centre-of-mass energy of 13 TeV. Generator uncertainties have been estimated using {\tt{Sherpa 2.2}} and are indicated by the orange band in the ratio panels. In addition, variations of the FxFx matching scale, $\mu_{Q}$, (shown in solid and dashed green) have been studied for the {\tt{MG5{\_}aMC@NLO+Py8}} setup. The statistical uncertainty component is indicated by the error bars. It can be seen that neither the PDF and scale variations estimated using {\tt{Sherpa 2.2}} nor the matching-scale variations estimated for the FxFx setup can cover the differences between the two predictions seen in the transition region, e.g. $\sqrt{d_{k}}\approx5$~GeV, between the soft and the perturbative regimes, where $\sqrt{d_{k}}$ is the associated splitting scale defined in Ref.~\cite{kTscales}.

\begin{figure}[t]
\centering
\subfigure[\label{shapeComp}]{\includegraphics[width=.48\textwidth]{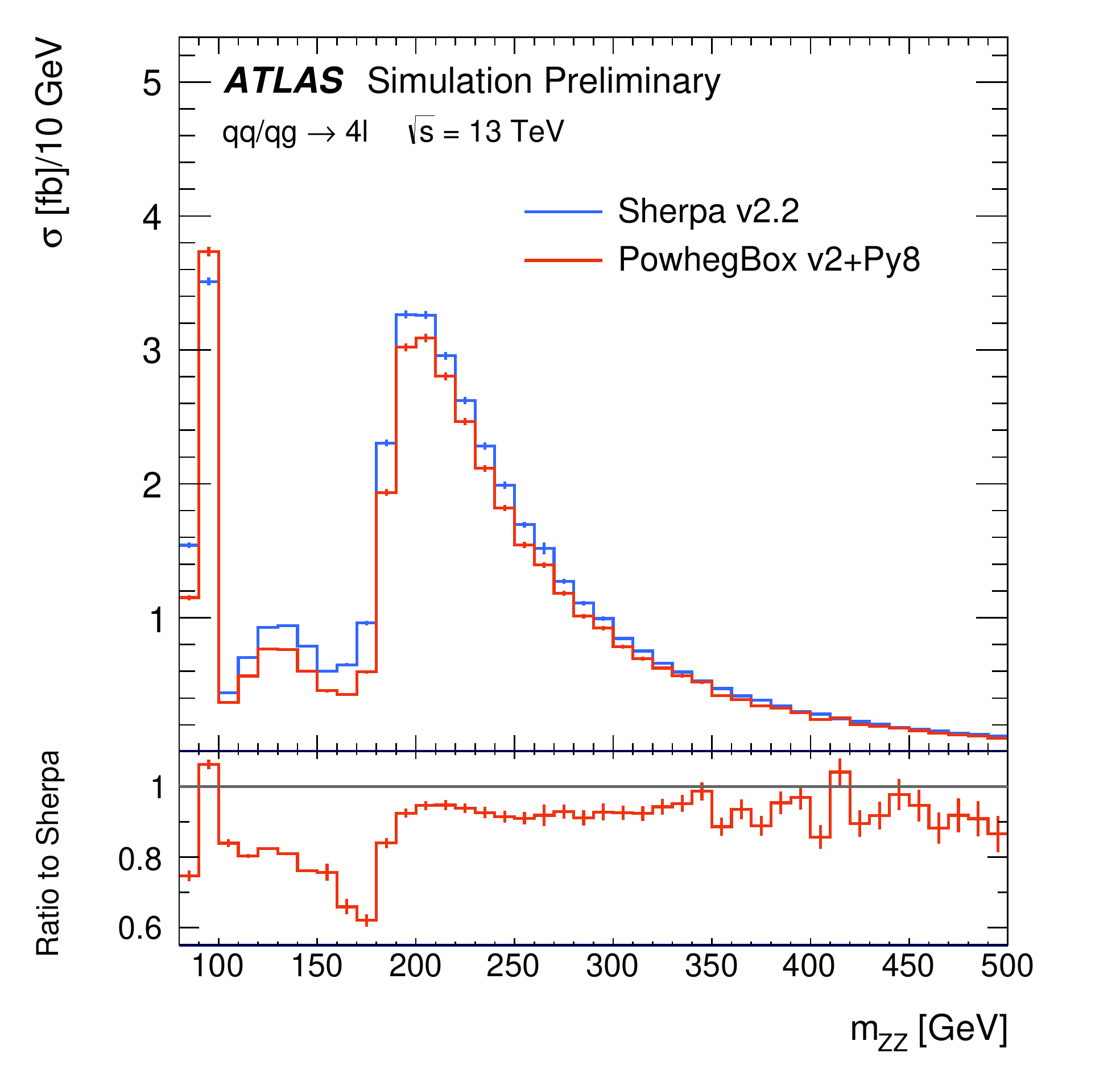}}
\subfigure[\label{EWcorr}]{\includegraphics[width=.48\textwidth]{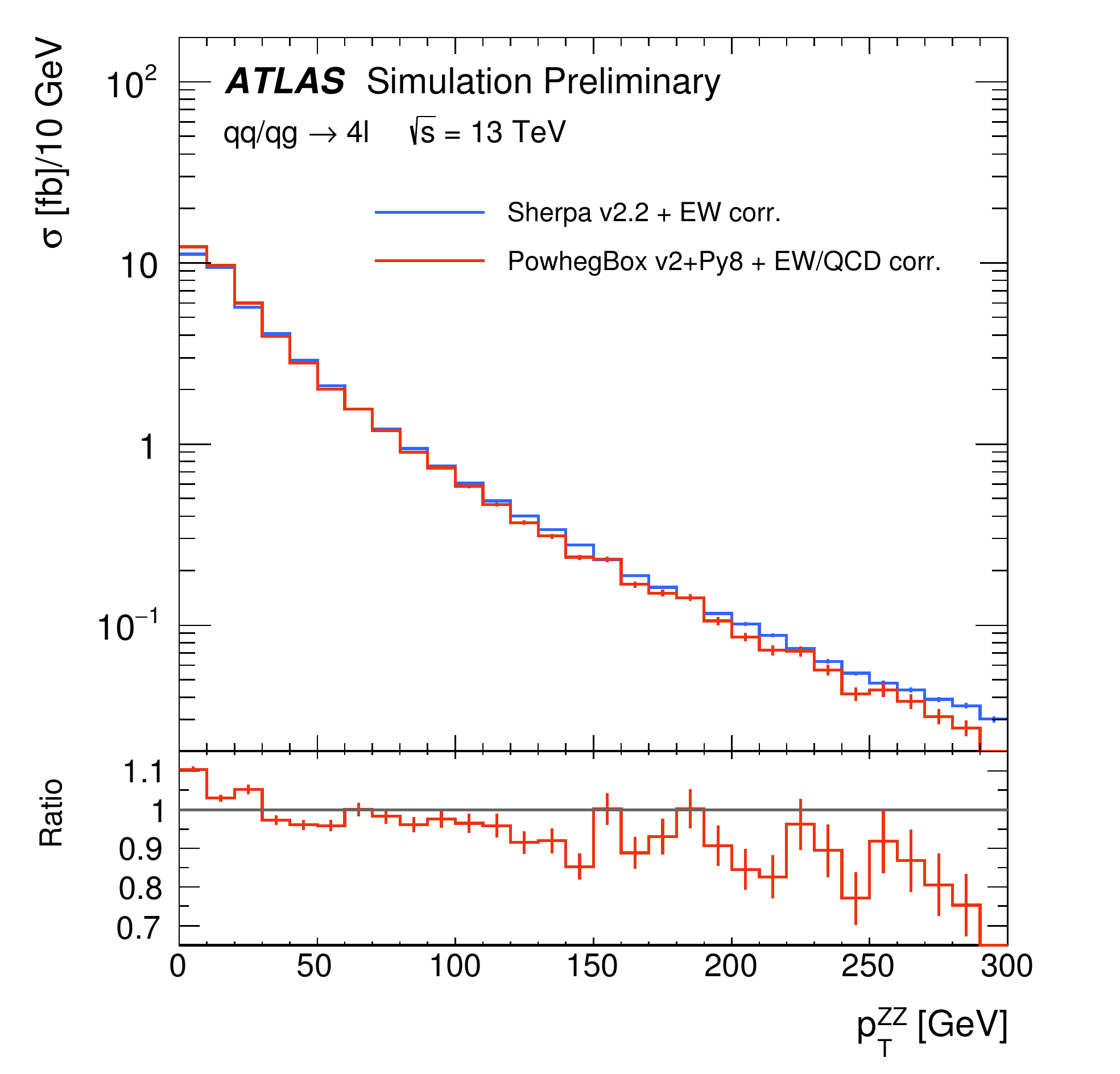}}
\caption{(a) Comparison of shapes ($d\sigma/dm_{ZZ}$) normalised to the samples' cross section and combining all four channels as predicted by {\tt{PowhegBox}} and {\tt{Sherpa}}; (b) Dfferential cross section predictions ($d\sigma/dp_{TT}$) of {\tt{Sherpa}} and {\tt{PowhegBox}} both corrected for higher-order electroweak effects and additional incorporation of QCD effects in {\tt{PowhegBox}}. These plots are taken from Ref.~\cite{multiboson}.}
\end{figure}

\paragraph{Multi-boson}
\subparagraph{Fully leptonic diboson processes}
The processes can be grouped according to the number of charged leptons, giving rise to the following final states: $4\ell$, $3\ell\nu$, $2\ell2\nu$, $\ell3\nu$ and $4\nu$. An overview of the accuracy achieved with the chosen generators is given in Table~1. A general shape comparison between {\tt{PowhegBox}} and {\tt{Sherpa v2.2}} is shown in Figure~\ref{shapeComp}.\\
\begin{table}[htbp]
\centering
\small
\begin{tabular*}{\textwidth}{ l c c c c c c }
\hline
 & & $VV+0j$ & $VV+1j$ & $VV+2j$ & $VV+3j$ & $VV+\geq 4j$ \\
\hline
\multirow{2}{*}{$ $}         & \tt{Sherpa v2.2}  & NLO & NLO & LO & LO & PS \\
                             & \tt{PowhegBox+PYTHIA8}/\tt{Herwig++}    & NLO &  LO & PS & PS & PS \\
                             & \tt{MadGraph5{\_}aMC@NLO+PYTHIA8}      & NLO &  NLO & LO & PS & PS \\

\hline
\end{tabular*}
\label{accuracies}
\caption{Overview of process accuracies for the chosen generators.}
\end{table}
The distributions are normalised to the samples' cross section and there is no reweighting applied. In Figure~\ref{shapeComp}, the four-lepton invariant mass, which is an observable quite insensitive to higher-order QCD effects, shows good agreement above the \textit{ZZ} threshold between the two generators. The deviations below this threshold can be related to differences in both the QCD and the electroweak shower which is fully based on {\tt{Pythia8}} in case of {\tt{PowhegBox}} while {\tt{Sherpa}} uses its own shower model. Higher-order electroweak corrections have been studied and it has been found that  they can also significantly affect several observables, particularly in the tails of the distributions. In Figure~\ref{EWcorr}, a comparison of the differential cross section ($d\sigma/dp^{ZZ}_\mathrm{T}$) predicted by {\tt{Sherpa}} and {\tt{PowhegBox}} is shown. While the predictions of both generators are similarly corrected for higher-order electroweak effects, there is an additional reweighting applied for {\tt{PowhegBox}} intended to bring it from NLO to approximately NNLO QCD accuracy.

\begin{figure}[tbp]
\begin{center}
\subfigure[]{\includegraphics[width=.48\textwidth]{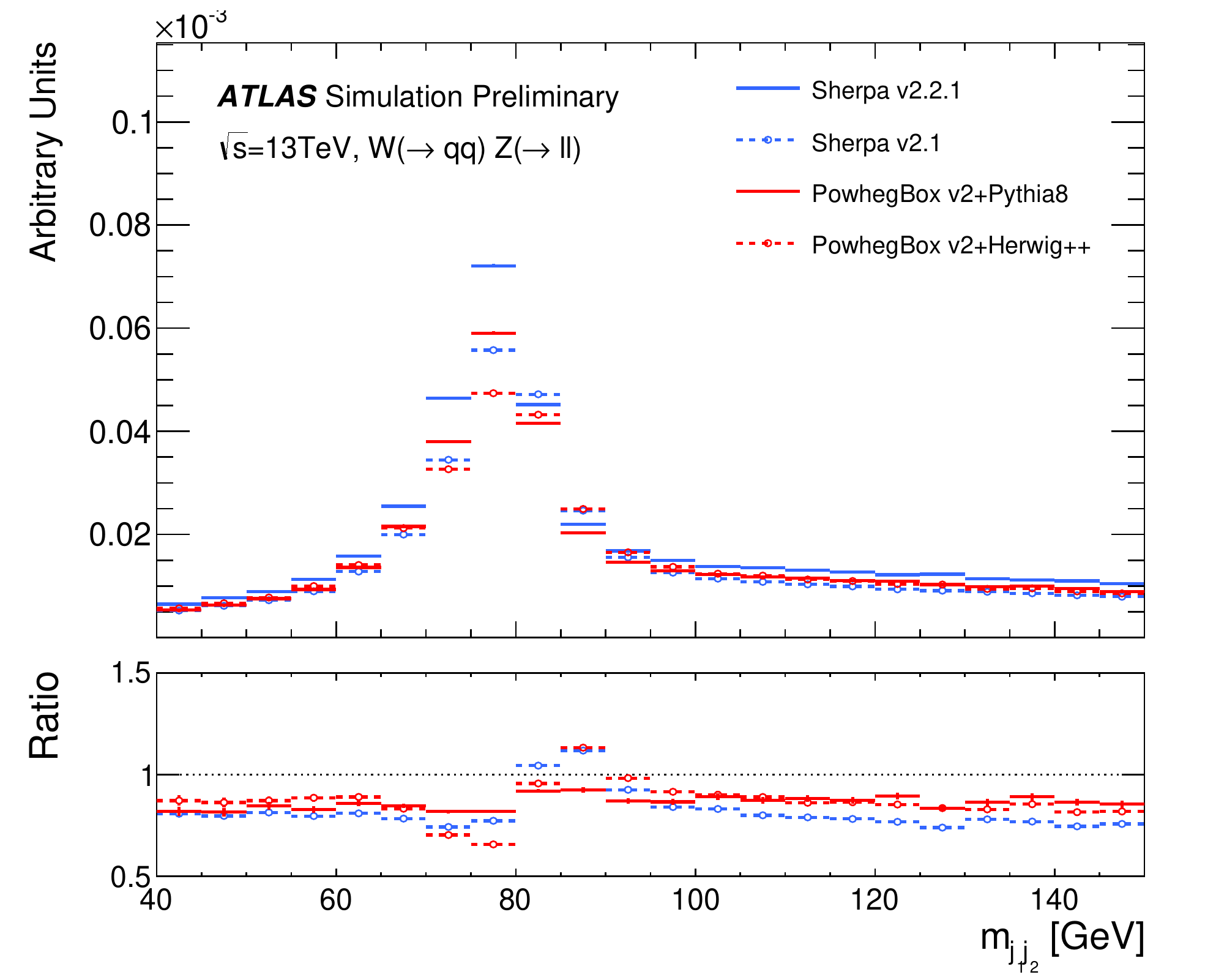}}
\subfigure[]{\includegraphics[width=.48\textwidth]{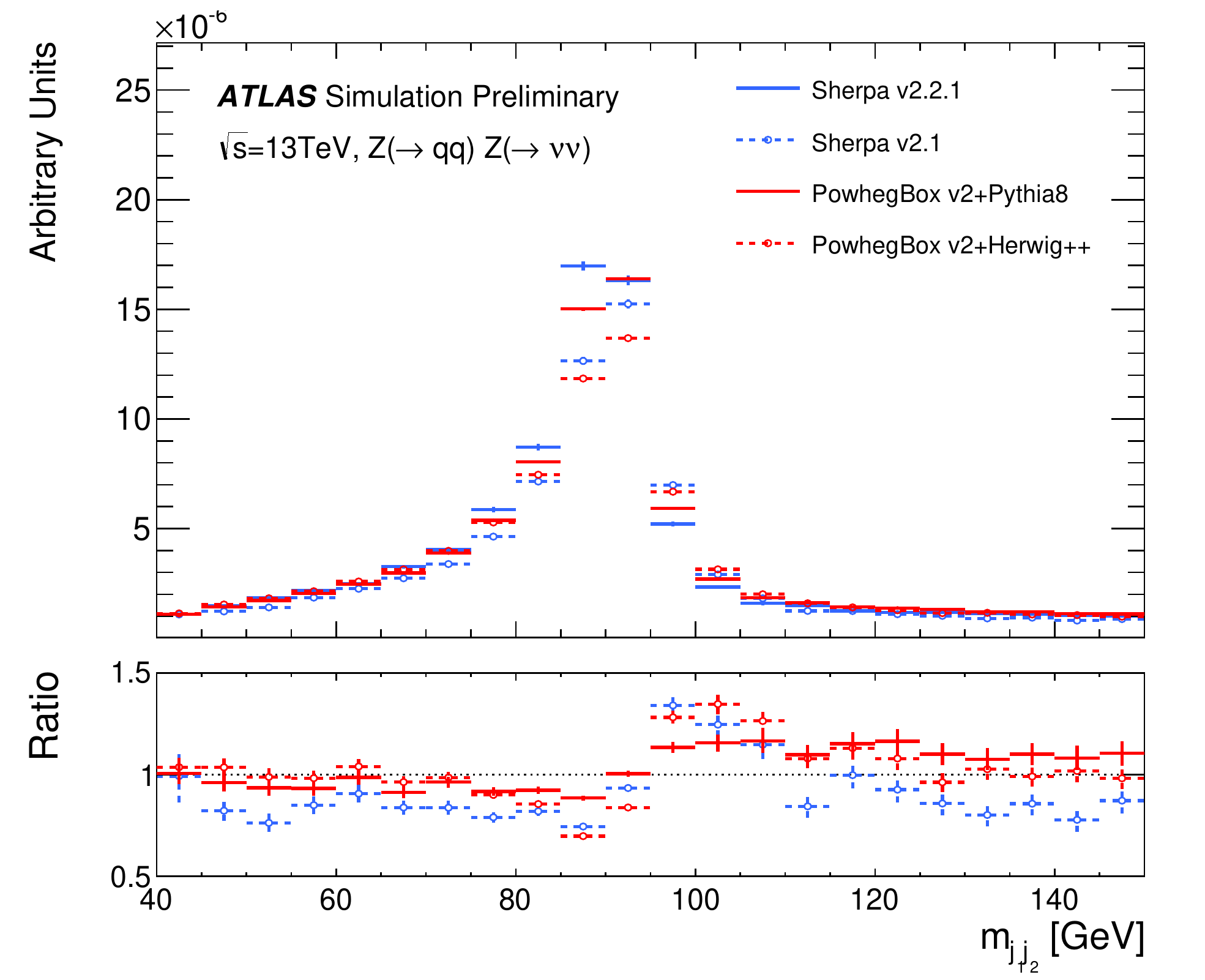}}
\end{center}
 \caption{Comparison of the invariant dijet mass $m(j_1,j_2)$ distribution in two different semileptonic final states at 13 TeV. The lower panel shows the ratio of each distribution with respect to the {\tt{Sherpa v2.2}} prediction. These plots are taken from Ref.~\cite{multiboson}.}
\label{fig:vv_mjj}
\end{figure}

\subparagraph{Semi leptonic diboson processes}
As regards the generators setup, an overview is given in Table~2. Figure~\ref{fig:vv_mjj} focuses on the comparison among different generators for the invariant mass of the reconstructed hadronically decaying boson, $m(j_1,j_2)$, for all diboson processes considered.
\begin{table}[htbp]
\centering
\label{accuracies2}
\small
\begin{tabular*}{0.9\textwidth} { l c c c c c c }
 \hline
  & & $VV+0j$ & $+1j$ & $+2j$ & $+3j$ & $+\geq 4j$ \\
 \hline
\multirow{3}{*}{$VV = WW, WZ$}        & {\tt{Sherpa v2.1.1}}  & NLO &  LO & LO & LO & PS \\
                                       & {\tt{Sherpa v2.2}}   & NLO & NLO & LO & LO & PS \\
                                       & {\tt{PowhegBox+Py8/Herwig++}}         & NLO &  LO & PS & PS & PS \\
 \hline
 \multirow{3}{*}{$VV = ZZ$}            & {\tt{Sherpa v2.1.1}}   & NLO & NLO & LO & LO & PS \\
                                       & {\tt{Sherpa v2.2}}   & NLO & NLO & LO & LO & PS \\
                                       & {\tt{PowhegBox+Py8/Herwig++}}         & NLO &  LO & PS & PS & PS \\
 \hline
\end{tabular*}
\caption{Overview of process accuracies for the chosen generators.}
\end{table}
It has been found that the {\tt{PowhegBox}} predictions are consistent with each other, indipendent of the choice from the program used to implement the showering and that they are, in general, softer than the {\tt{Sherpa v2.2}} ones. A similar conclusion can be drawn if analysing the $WZ(\rightarrow\ell\ell)$ and the $WZ(\rightarrow\nu\nu)$ channels.

\begin{figure}[t]
\centering
\subfigure[]{\includegraphics[width=.48\textwidth]{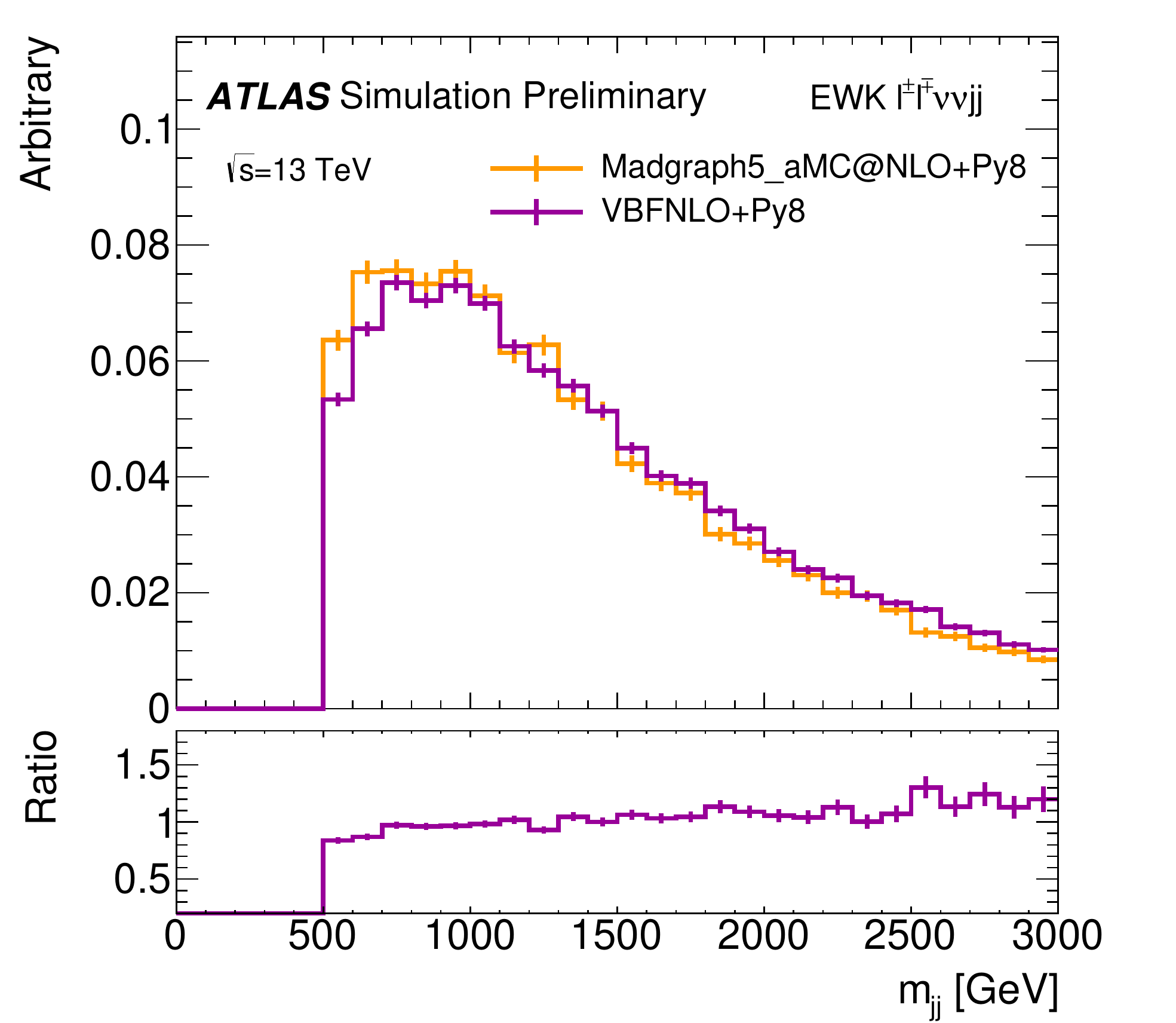}}
\subfigure[]{\includegraphics[width=.48\textwidth]{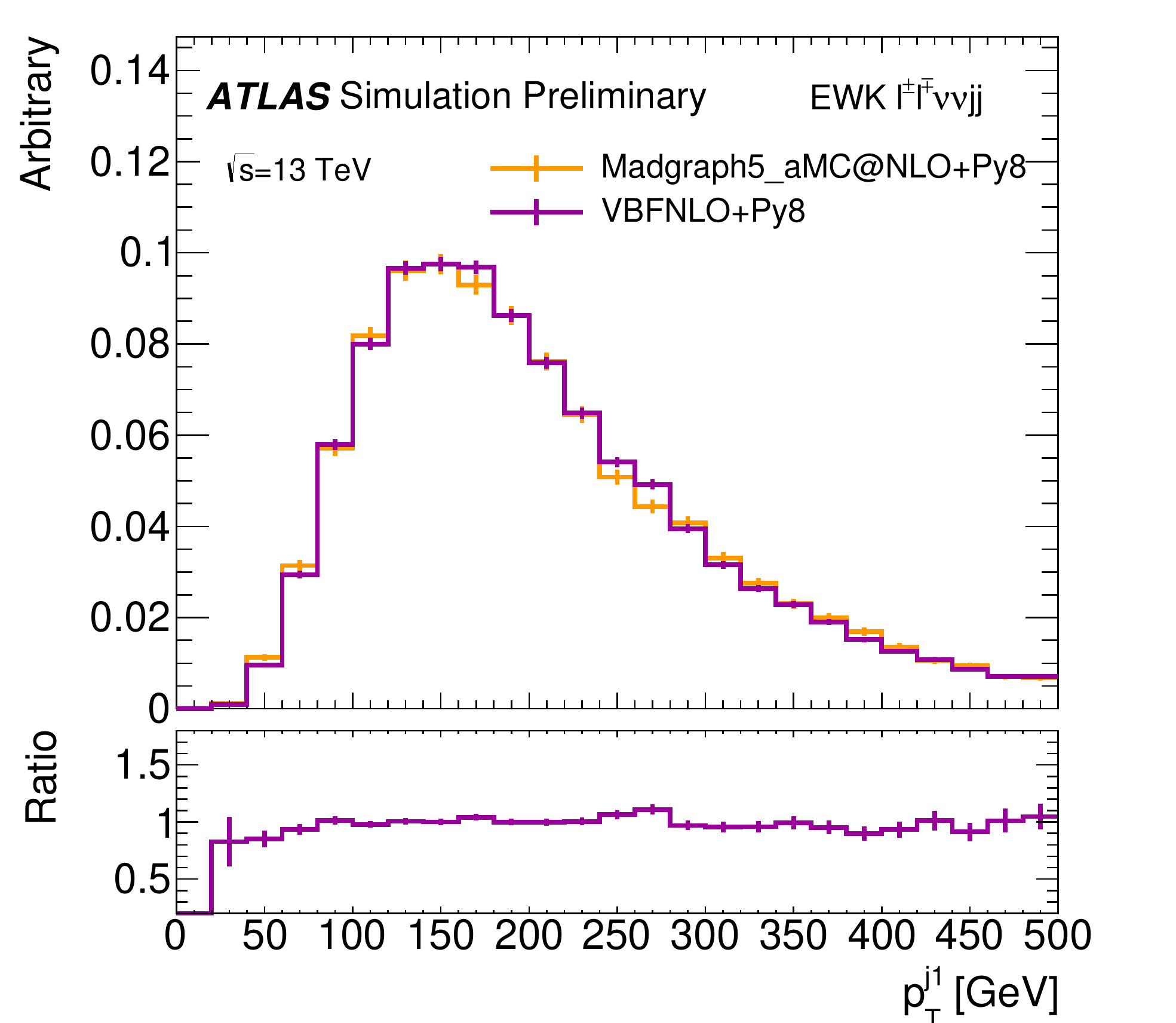}}
\caption{Comparison of predicted kinematic distributions from {\tt{MG5{\_}aMC@NLO}} and {\tt{VBFNLO}}, both showered with {\tt{Pythia8}}, for $m_{jj}$ (left) and leading jet $p_\mathrm{T}$ (right). These plots are taken from Ref.~\cite{multiboson}.}
\label{VBFNLO}
\end{figure}

\begin{figure}[t]
\centering
\subfigure[\label{EW_Sherpa}]{\includegraphics[width=.48\textwidth]{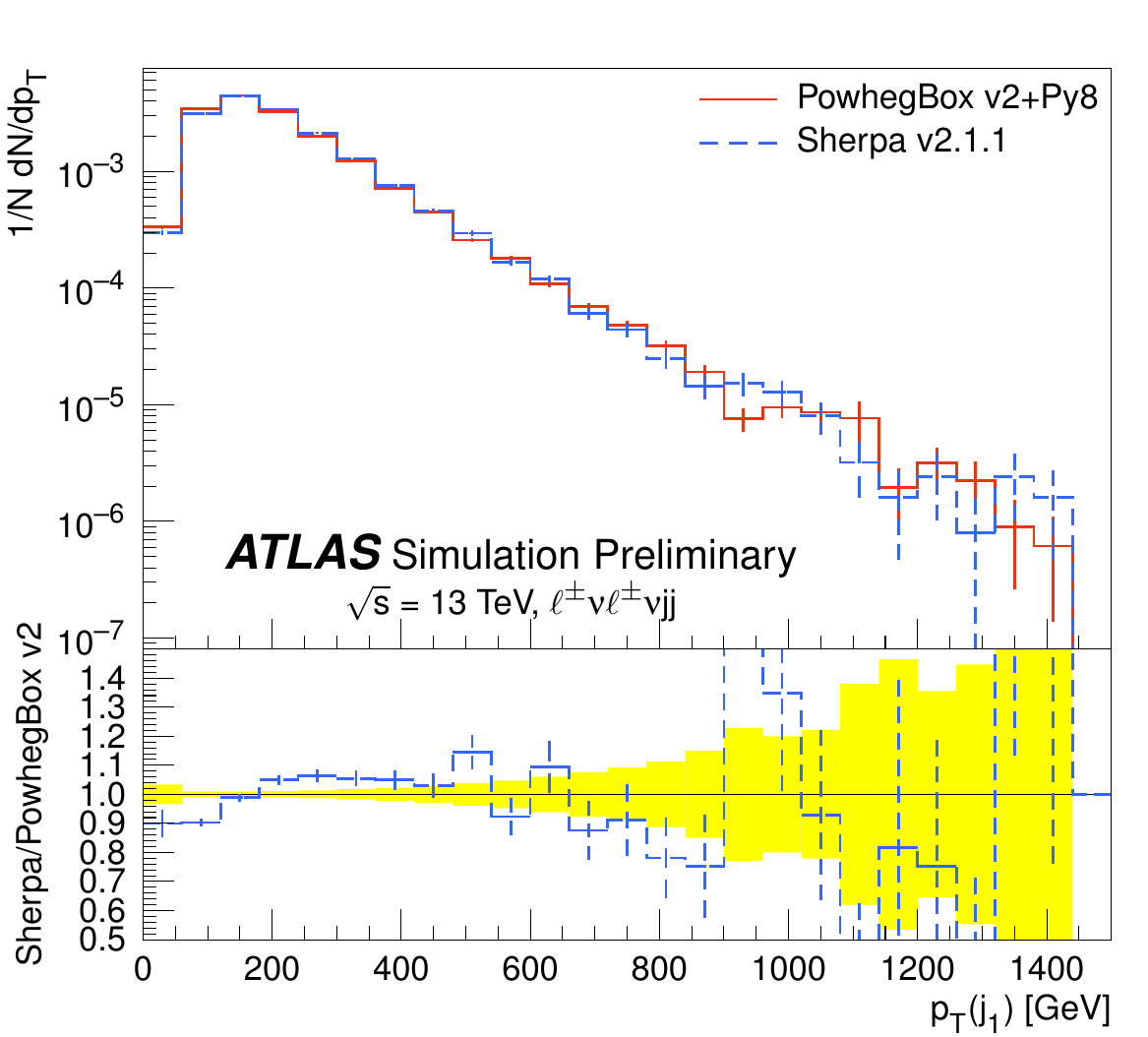}}
\subfigure[\label{SysUnc}]{\includegraphics[width=.48\textwidth]{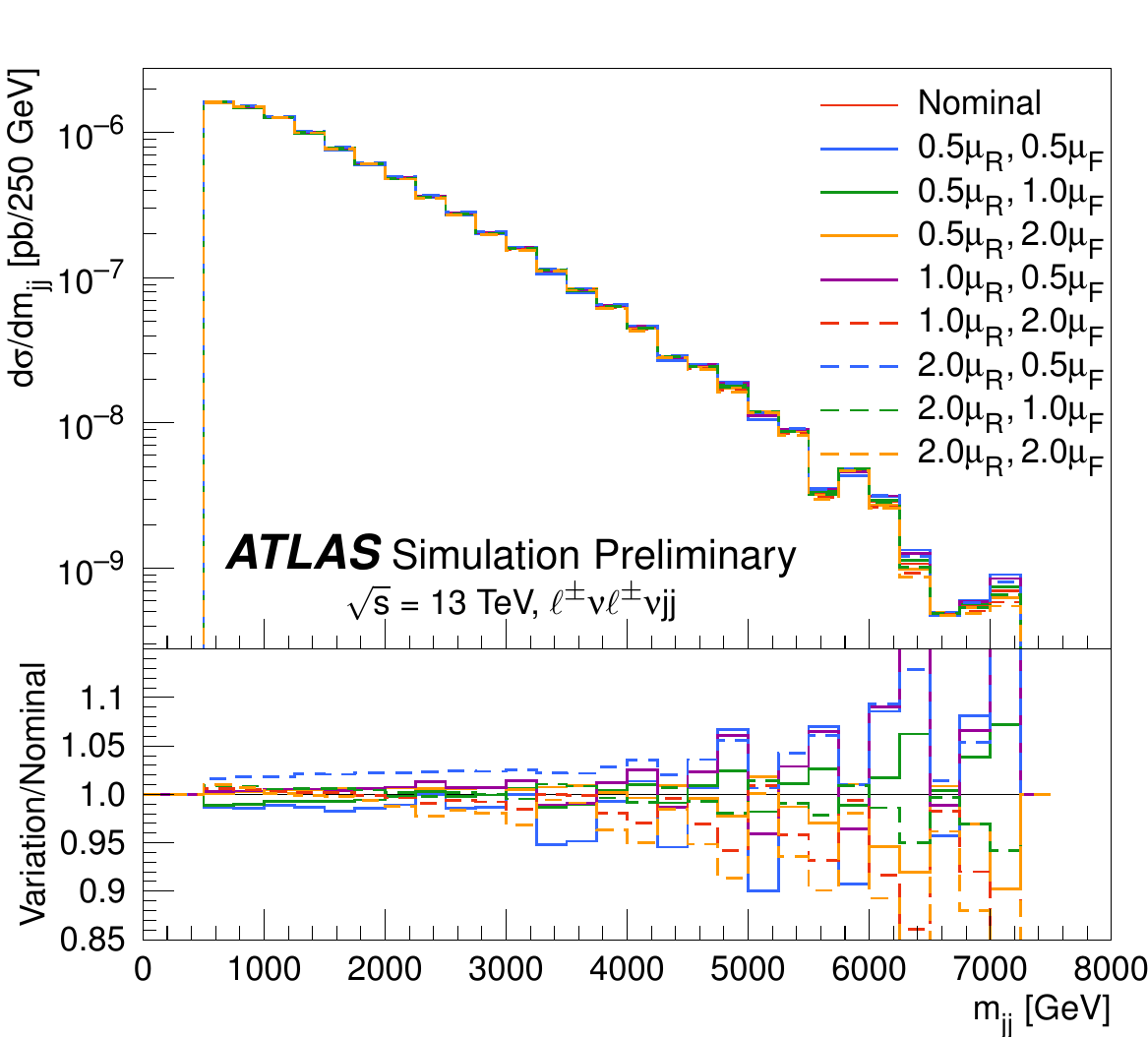}}
\caption{(a) Comparison between the leading jet $p_\mathrm{T}$ distribution predicted by  {\tt{PowhegBox}} and {\tt{Sherpa}}. The yellow band represents the statistical uncertainty of the {\tt{PowhegBox}} sample; (b) Impact of scale variations on the dijet invariant mass $m_{jj}$. These plots are taken from Ref.~\cite{multiboson}.}
\end{figure}

\subparagraph{Electroweak diboson production with jets}
Electroweak diboson production with at least two jets includes vector boson scattering (VBS) diagrams, where the two "tagging" jets recoil against the (heavy) gauge bosons. The resulting leptonic final states include the $4 \ell jj$, as well as the $2 \ell 2\nu jj$ final states, where the two lepton charges can be the same or opposite sign. An overview of the accuracy achieved with the chosen generators is given in Table~3.
\begin{table}[htbp]
\centering
\caption{Accuracies of the chosen generators for the listed electroweak processes.}
\label{tab:ewk-vv-accuracies}
\begin{tabular*}{0.95\textwidth} { l c c c c }
 \hline
  & & $VV+2j$ & $VV+3j$ & $VV+\geq 4j$ \\
 \hline
 \multirow{2}{*}{$VVjj = \ell^{\pm}\ell^{\mp} 2\nu jj$} 
                                                    &{\tt{VBFNLO+Py8}} & LO & PS & PS \\
                                                  & {\tt{MG5{\_}aMC@NLO+Py8}} & LO & PS & PS \\	
  \hline
 \multirow{2}{*}{$VVjj = \ell^{\pm}\ell^{\pm} 2\nu jj$} & {\tt{Sherpa v2.1.1}} & LO & PS & PS \\
                                                   & {\tt{PowhegBox+Py8}} & NLO & LO & PS \\
  \hline
\end{tabular*}
\end{table}
A comparison of predicted kinematic distributions from {\tt{MG5{\_}aMC@NLO}} and {\tt{VBFNLO}}, both showered with {\tt{Pythia8}} is shown in Figure~\ref{VBFNLO}: these distributions are found to be similar.
Furthermore, the comparison between the leading jet $p_\mathrm{T}$ distribution predicted by {\tt{PowhegBox}} and {\tt{Sherpa}} can be found in Figure~\ref{EW_Sherpa}; {\tt{PowhegBox}} prediction is overall in agreement within uncertainties with what is seen in {\tt{Sherpa}}.
The systematic variations for events generated with {\tt{PowhegBox}} are derived using the {\tt{PowhegBox}} internal reweighting scheme. The resulting weights include renormalisation scale and factorisation scale. There are eight scale variations corresponding to factors of 1/2 or 2 applied independently to the
renormalisation and factorisation scales, as shown in Figure~\ref{SysUnc} for the dijet system invariant mass distribution. The impact of scale variation is moderate (2-3\%) for $m_{j_{1},j_{2}}$ up to 4 TeV, and then it becomes larger, rising up to 10-15\% for $m_{j_{1},j_{2}}\approx  7$ TeV.

\end{document}